\documentclass[fleqn,usenatbib]{mnras}

\usepackage{newtxtext,newtxmath}
\usepackage[T1]{fontenc}
\usepackage{gensymb}  %added by Jay
\DeclareRobustCommand{\VAN}[3]{#2}
\let\VANthebibliography\thebibliography
\def\thebibliography{\DeclareRobustCommand{\VAN}[3]{##3}\VANthebibliography}

\usepackage{graphicx}	% Including figure files
\usepackage{amsmath}	% Advanced maths commands
\usepackage{float}  %added by Jay
\title[eMSTO in the open cluster NGC 2355]{Origin of extended Main Sequence Turn Off in open cluster NGC 2355}

\author[Jayanand Maurya et al.]{
Jayanand Maurya,$^{1}$\thanks{E-mail: maurya.jayanand@gmail.com}
M. R. Samal,$^{2}$
Louis Amard,$^{3}$
Yu Zhang,$^{1,4}$ 
Hubiao Niu,$^{1}$
\newauthor{Sang Chul Kim,$^{5,6}$ Y. C. Joshi,$^{7}$ and B. Kumar$^{7}$}
\\
$^{1}$Xinjiang Astronomical Observatory, Chinese Academy of Sciences, No. 150, Science 1 Street, Urumqi, Xinjiang 830011, PR China\\
$^{2}$Astronomy \& Astrophysics Division, Physical Research Laboratory, Ahmedabad, 380009, India\\
$^{3}$Universit\'e Paris-Saclay, Universit\'e Paris Cit\'e, CEA, CNRS, AIM, F-91191, Gif-sur-Yvette, France\\
$^{4}$School of Astronomy and Space Science, University of Chinese Academy of Sciences, Beijing 100049, People's Republic of China\\
$^{5}$Korea Astronomy and Space Science Institute, 776, Daedeokdae-ro, Yuseong-gu, Daejeon 34055, Republic of Korea\\
$^{6}$Korea University of Science and Technology (UST), Daejeon 34113, Republic of Korea\\
$^{7}$Aryabhatta Research Institute of Observational Sciences (ARIES), Manora peak, Nainital 263002, India
}

\date{Accepted 2024 June 24. Received 2024 June 14; in original form 2023 September 27}

\pubyear{2024}

\begin{document}
\label{firstpage}
\pagerange{\pageref{firstpage}--\pageref{lastpage}}
\maketitle

\begin{abstract}
The presence of extended Main Sequence Turn-Off (eMSTO) in the open clusters has been attributed to various factors, such as spread in rotation rates, binary stars, and dust-like extinction from stellar excretion discs. We present a comprehensive analysis of the eMSTO in the open cluster NGC 2355. Using spectra from the \textit{Gaia}-ESO archives, we find that the stars in the red part of the eMSTO have a higher mean \textit{v} sin \textit{i} value of 135.3$\pm$4.6 km s$^{-1}$ compared to the stars in the blue part that have an average \textit{v} sin \textit{i} equal to 81.3$\pm$5.6 km s$^{-1}$. This suggests that the eMSTO in NGC 2355 is possibly caused by the spread in rotation rates of stars. We do not find any substantial evidence of the dust-like extinction from the eMSTO stars using ultraviolet data from the \textit{Swift} survey. The estimated synchronization time for low mass ratio close binaries in the blue part of the eMSTO suggests that they would be mostly slow-rotating if present. However, the stars in the blue part of the eMSTO are preferentially located in the outer region of the cluster indicating that they may lack low mass ratio close binaries. The spread in rotation rates of eMSTO stars in NGC 2355 is most likely caused by the star-disc interaction mechanism. The stars in the lower main sequence beyond the eMSTO region of NGC 2355 are slow-rotating (mean \textit{v} sin \textit{i} = 26.5$\pm$1.3 km s$^{-1}$) possibly due to the magnetic braking of their rotations.
\end{abstract}


%%%%%%%%%%%%%%%%%%%%%%%%%%%%%%%%%%%%%%%%%%%%%%%%%%

%%%%%%%%%%%%%%%%% BODY OF PAPER %%%%%%%%%%%%%%%%%%

\section{Introduction}

The unusual spread near the turn-off of the main sequence (MS) in the colour-magnitude diagram of the Galactic open clusters has drawn the attention of astronomers recently \citep{2018ApJ...863L..33M}.  This spread in the upper MS is called extended main sequence turn-off i.e. eMSTO. Though recent in detection, the eMSTOs are now commonly found in the Galactic open clusters \citep{2018ApJ...869..139C,2019AJ....158...35S,2021AJ....162...64M}. It has been found that the inferred apparent age spread from the eMSTO's width is correlated to the cluster age \citep{2015MNRAS.453.2070N,2018MNRAS.480.3739B}. This indicates that the presence of the eMSTOs is related to stellar evolution rather than star formation. \citet{2015MNRAS.453.2070N} investigate the origin of the eMSTO in the clusters and predict that the stellar rotation can cause broadening of the upper MS of the clusters which was found to be true for observed young and intermediate-age clusters \citep{2018MNRAS.480.3739B,2019ApJ...876..113S}. The fast rotational velocity causes a decrease in the self-gravity of the star which manifests in the reduced stellar surface temperature, a phenomenon known as gravity darkening. Therefore, the fast-rotating stars appear redder on the colour-magnitude diagram (CMD) due to the gravity-darkening effect \citep{1924MNRAS..84..665V}. However, gravity darkening affects the stellar equators more significantly than the poles of the stars so the observed colour of fast-rotating stars also depends on the inclination angle. Additionally, the rotational mixing in the fast-rotating stars can enhance their core size and lifetime on MS. These stars will appear younger than their non-rotating counterparts \citep{2000ARA&A..38..143M}. 

Although the spread in rotation rates of stars is the widely accepted reason behind the origin of eMSTO in the open clusters, there are studies suggesting other possible mechanisms that could contribute to the presence of the eMSTOs in the open clusters \citep{2019MNRAS.490.2414P,2022MNRAS.512.3992C}. The differential reddening was the cause behind the presence of eMSTO in cluster Stock 2 instead of the rotational velocity \citep{2021A&A...656A.149A}. However, the open clusters hosting eMSTOs explored by \citet{2018ApJ...869..139C} do not show any significant differential reddening. Similarly, using simple stellar populations models \citet{2022MNRAS.512.3992C} notify the contribution of the binary stars in the eMSTO morphologies of the open clusters. Recently, \citet{2023MNRAS.521.4462D} propounded the idea that dust-like extinction from the circumstellar excretion disc of the fast-rotating stars may lead to the presence of the eMSTO in the star clusters. Thus, the exact reason behind the origin of the presence of eMSTO in the Galactic open clusters is still a debated topic.

The possible mechanism causing the spread in rotation rates of the stars with similar masses belonging to the eMSTO is still debated. \citet{2015MNRAS.453.2637D,2017NatAs...1E.186D} proposed that all the stars were initially fast-rotating stars before tidal torque in binary stars, causing them to become slow rotators. The finding that a few star clusters have comparable binary fractions in the fast and slow-rotating stars samples seems not to support this hypothesis \citep{2020MNRAS.492.2177K,2021MNRAS.508.2302K}. A different explanation for the spread in the rotation rate of MS stars having masses above $\sim$1.5 M$_{\odot}$ links this to the bimodal stellar rotation distribution in the pre-main sequence (PMS) low-mass ($\la$2 M$_{\odot}$) stars \citep{2020MNRAS.495.1978B}. This bimodal stellar rotation in the PMS stars depends on the interaction of stars with their circumstellar discs during PMS lifetimes. Another interesting mechanism happening in very young clusters suggests that the fast rotation is caused by disc accretion and slow rotation is due to the binary merger in the stars \citep{2022NatAs...6..480W}. So, theories providing possible physical mechanisms leading to the spread in rotation rates of the eMSTO stars are still emerging. Therefore, studying the origin of the eMSTOs in open star clusters needs keen attention to develop a better understanding of this physical phenomenon. 

In this work, we carefully investigate the origin of the eMSTO in open cluster NGC 2355. The open cluster NGC 2355 is an intermediate-age open cluster having an age of 1 Gyr \citep{2020A&A...640A...1C}. This cluster is located at a distance of 1941 pc with Galactic longitude and latitude of 203.370 and 11.813 degrees, respectively \citep{2020A&A...640A...1C}. The cluster is situated in the anticenter direction in the Gemini constellation at the Galactocentric distance of 10.1 kpc near the region where the Milky Way metallicity gradient flattens \citep{2020A&A...640A...1C,2022AJ....164...85M}. The mean metallicity of the cluster is estimated to be [Fe/H] = -0.09 by \citet{2021A&A...651A..84M} using medium resolution \textit{Gaia}-ESO  UVES (R$\sim$47000) and GIRAFFE HR15N  (R$\sim$19000) spectra. This [Fe/H] metallicity translates to Z = 0.0163 in the mass fraction form of metallicity using the relation provided by \citet{1994A&AS..106..275B}. The broadening near the MS turn-off of NGC 2355 was noticed by \citet{2019AJ....158...35S} using \textit{Swift}  near-ultraviolet (1700-3000 Å) data without further spectroscopic study. The open cluster NGC 2355 is also studied for stellar variability by \citet{2022AJ....164...40W} using optical V band data. A total of 15 members are variable stars in the study. 

The present study is organized as follows. In Section 2, we present the source and specifications of the data used in the current analysis. We describe member identification, cluster general properties, and demography and physical properties of the eMSTO stars in Section 3. In Section 4, we explore the different reasons that could explain the origin of the eMSTO. Finally, we conclude in Section 5.
\section{Data}
We utilized photometric and astrometric data provided by \textit{Gaia} Data Release 3 (DR3) \citep{2016A&A...595A...1G,2022arXiv220800211G} for the present study.   The \textit{Gaia} DR3 archive provides data with photometric uncertainty of 0.0003, 0.001, and 0.006 mag for the G band data up to G<13, G=17, and G=20 mag, respectively. The photometric uncertainties for the G$_{\rm BP}$ band  are $\sim$0.0009, 0.012, and 0.108 mag up to  G<13,  G=17, and G=20 mag, respectively.  The G$_{\rm RP}$ band data have  0.0006, 0.006, and 0.052 mag photometric uncertainty for G<13,  G=17, and G=20 mag, respectively. The proper motions provided by \textit{Gaia} DR3 have median uncertainty of $\sim$0.03, 0.07, and 0.5 mas/yr for G<15, G=17, and G=20 mag, respectively. The median uncertainties in parallax are $\sim$0.03 mas for G<15, 0.07 mas for G=17, and 0.5 mas for G=20 mag. We used \textit{Gaia} DR3 astrometric data to identify member stars and calculate the distance of the cluster NGC 2355.

The ultraviolet (UV) data for NGC 2355 in the uvw2 (1928 $\AA$), uvm2 (2246 $\AA$), and uvw1 (2600 $\AA$) bands of the \textit{Swift} survey provided by \citet{2019AJ....158...35S} is used in the present study. This data set is part of the Swift/Ultraviolet-Optical Telescope Stars Survey and includes near-ultraviolet photometric data of 103 Galactic open clusters. We used this data to create the ultraviolet CMD for NGC 2355.

We used medium and high-resolution spectra available at the European Southern Observatory (ESO) archive \footnote{http://archive.eso.org/scienceportal/home}. The medium-resolution and high-resolution spectra were observed using the multi-object GIRAFFE and UVES spectrographs installed on the ESO Very Large Telescope (VLT). The archived spectra were observed under programme 197.B-1074 (PI: GILMORE, GERARD). The resolution for medium-resolution GIRAFFE spectra was R = $\frac{\lambda}{\Delta \lambda}$ $\sim$ 19,200 for the wavelength range $\sim$644 to $\sim$680 nm. The high-resolution spectra from the UVES spectrograph have a resolution of R $\sim$ 51,000 in the wavelength range $\sim$582-683 nm. These \textit{Gaia}-ESO spectra were used to estimate the projected rotational velocity of the stars in this paper.

\section{Membership and Physical Properties}

\subsection{Identification of the member stars}

\begin{figure}
	\includegraphics[width= 8.4 cm]{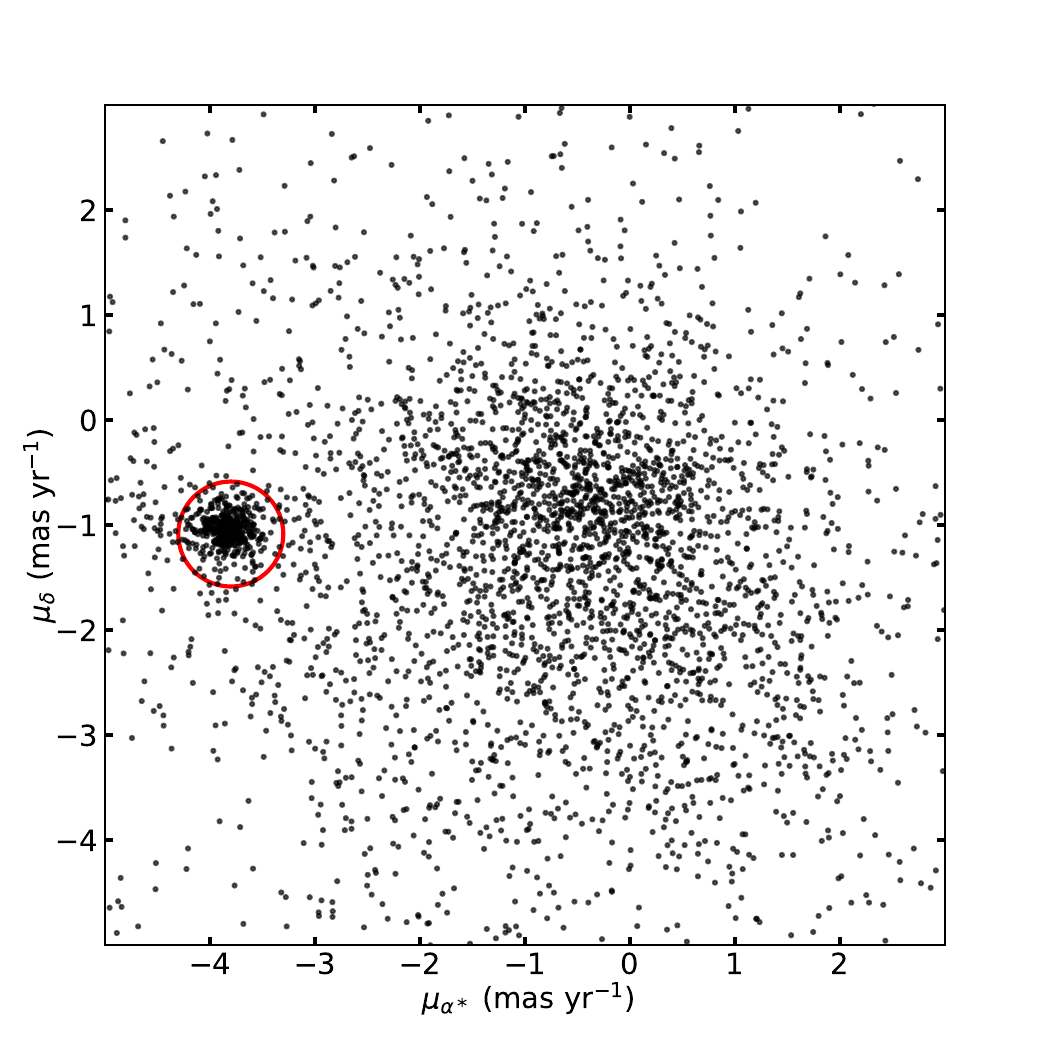}
    \caption{Plot of vector-point diagram for the cluster NGC 2355. The points enclosed by a red circle represent probable cluster members.}
    \label{vpd}
\end{figure}

The identification of member stars is necessary to derive conclusions about the origin of the eMSTO stars. For this purpose, we used very precise astrometric data from \textit{Gaia} DR3 archives. We downloaded all the sources from \textit{Gaia} DR3 data within a radius of 0.261$\degree$ around the cluster center with right ascension (RA) = 07:16:59.3 and declination (Dec.) = +13:46:19.2 of epoch=J2000 given by \citet{2020A&A...640A...1C}. This radius is equal to the tidal radius of the cluster estimated by \citet{2022AJ....164...54Z}. We present a vector-point diagram (VPD) of the NGC 2355 region showing the proper motion plane in Figure~\ref{vpd}. We noticed a conspicuous over-density in the VPD of NGC 2355 at ($\mu_{\alpha *}$ = -3.802 mas yr$^{-1}$; $\mu_{\delta}$ = -1.086 mas yr$^{-1}$) as shown Figure~\ref{vpd}. The stars lying within a circle of radius 0.6  mas yr$^{-1}$ centered at this point were chosen as potential member stars. This radius is estimated through the radial density profile of stars in the proper motion plane as described in our previous studies \citep{2020MNRAS.492.3602J,2020MNRAS.495.2496M}. The radius is the radial distance from the center where the number density of potential members starts merging into the number density of the field stars in the proper motion plane. To quantify the membership of stars, we assigned membership probability to the stars calculated through a statistical method utilizing proper motions as described in \citet{2020MNRAS.494.4713M}. The stars with membership probabilities above 60$\%$ and parallaxes, $\varpi$, within 3$\sigma$ standard deviation of the mean $\varpi$ of the potential member stars were chosen as cluster members. Through this process, we obtained 411 member stars in cluster NGC 2355. 

\subsection{Extended Main Sequence Turn-Off in the cluster}\label{eMSTO}
\begin{figure}
	\includegraphics[width= 8.4 cm]{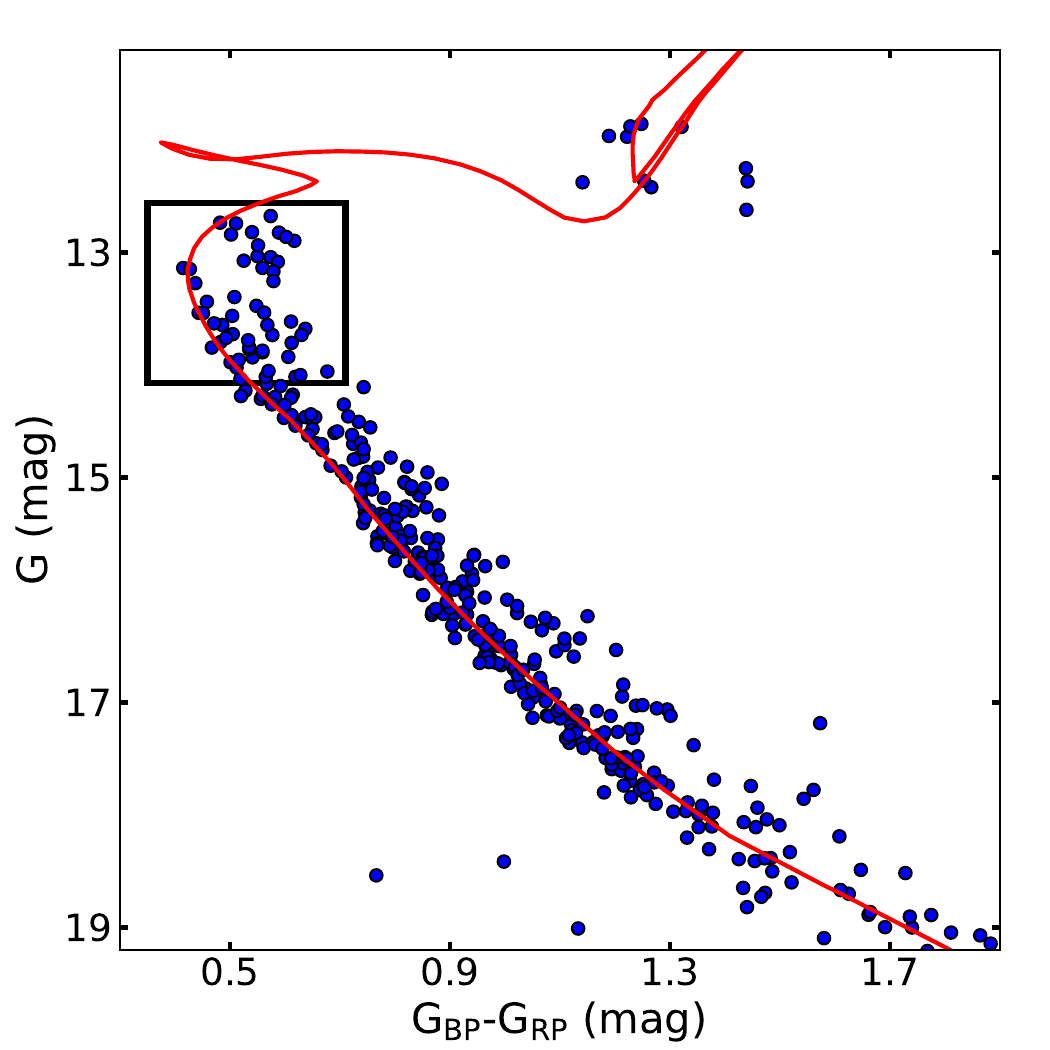}
    \caption{Plot of fitted \citet{2017ApJ...835...77M} isochrones on G/(G$_{\rm BP}$ - G$_{\rm RP}$) colour-magnitude diagram of NGC 2355. The best-fit isochrone shown by the red continuous curve corresponds to the age of 1.0 Gyr. The extended main sequence turn-off region is enclosed within the black rectangle drawn in the upper main sequence part of the CMD.}
    \label{age}
\end{figure}

\begin{figure}
	\includegraphics[width= 8.4 cm]{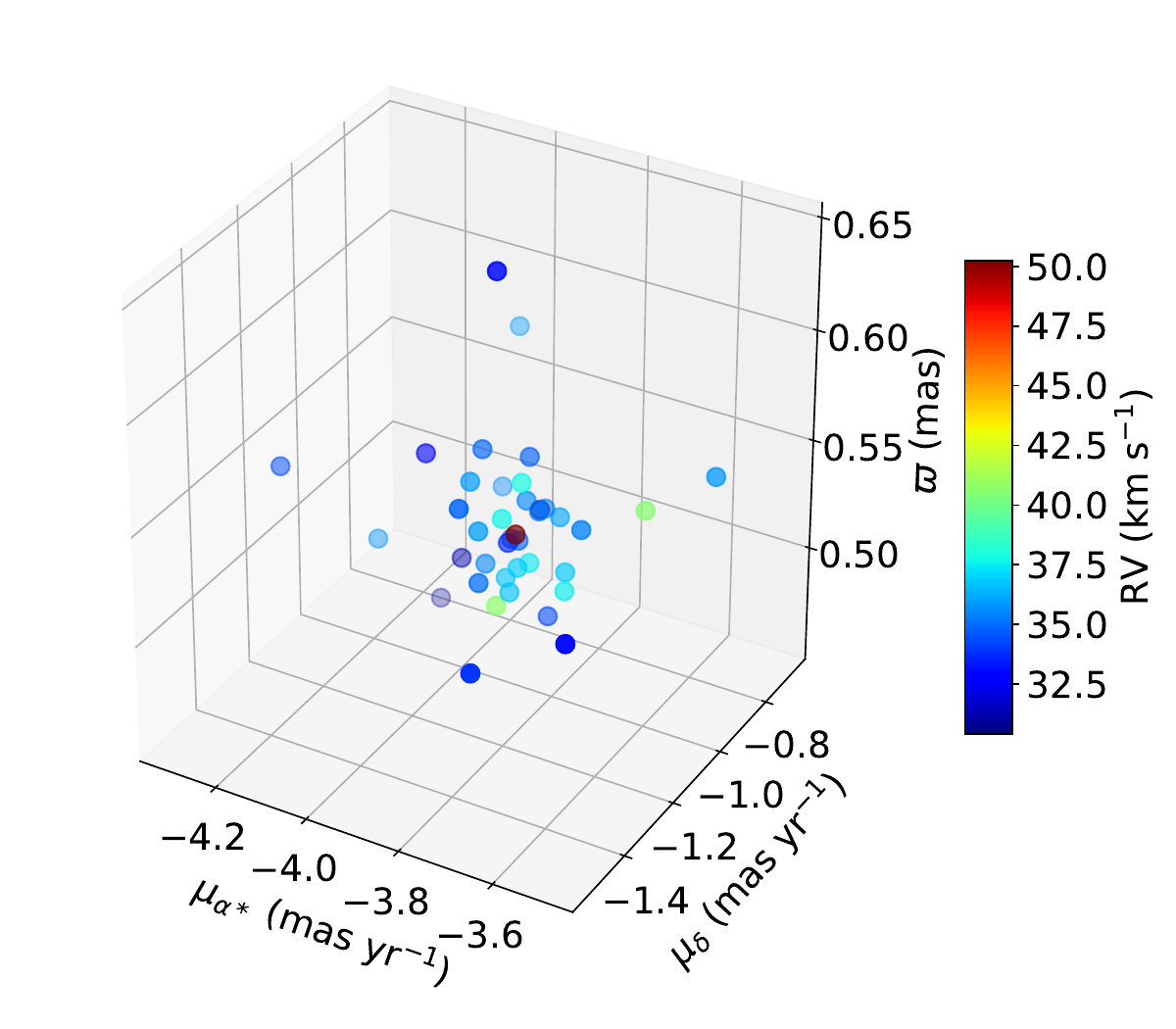}
    \caption{Plot for 3D kinematics of stars belonging to extended main sequence turn-off of the cluster NGC 2355. The points representing proper motions and parallax of stars are colour-coded with their radial velocities.}
    \label{3d_rv}
\end{figure}

We notice an unusual broadening of the upper MS, i.e. the eMSTO, in the CMD of NGC 2355, as visible in Figure~\ref{age}. We defined the extended main sequence turn-off region in the colour-magnitude diagram of the cluster NGC 2355 as the rectangular region with G$_{\rm BP}$ - G$_{\rm RP}$ colour values between 0.35 to 0.70 mag and G band magnitude less than 14.16 mag. We analyzed the 3D kinematics of the stars belonging to the eMSTO region which is illustrated in Figure~\ref{3d_rv}. The radial velocity (RV) of 39 out of 54 eMSTO stars were available in \citet{2022A&A...666A.121R} derived from the spectroscopic data from the \textit{Gaia}-ESO survey. We found all these 39 eMSTO stars except one share approximately similar RV with a mean RV value of 35.4$\pm$0.4 km s$^{-1}$ and a standard deviation of 2.1 km s$^{-1}$. The remaining eMSTO star with ID 38 is reported to have an RV value of 50.2 $\pm$ 0.3 km s$^{-1}$, which may be a binary star. The proper motions, parallaxes, and RV of 53 out of 54 eMSTO stars suggest that these stars are profound members of the cluster NGC 2355.

The stars in the eMSTO region have magnitudes G$\leq$14.16, G$_{\rm BP}$ $\leq$ 14.34, and   G$_{\rm RP}$ $\leq$ 13.80, respectively. The uncertainties for these magnitude ranges are estimated to be 0.0002,  0.0012, and 0.0010 mag, respectively, using the tool provided by \textit{Gaia} DPAC \citep{2021A&A...649A...3R}. The uncertainty in the (G$_{\rm BP}$ - G$_{\rm RP}$) colour would be 0.002 mag compared to the observed spread of 0.222 mag in the (G$_{\rm BP}$ - G$_{\rm RP}$) colour for the eMSTO stars. 
\begin{figure}
\includegraphics[width= 8.4 cm]{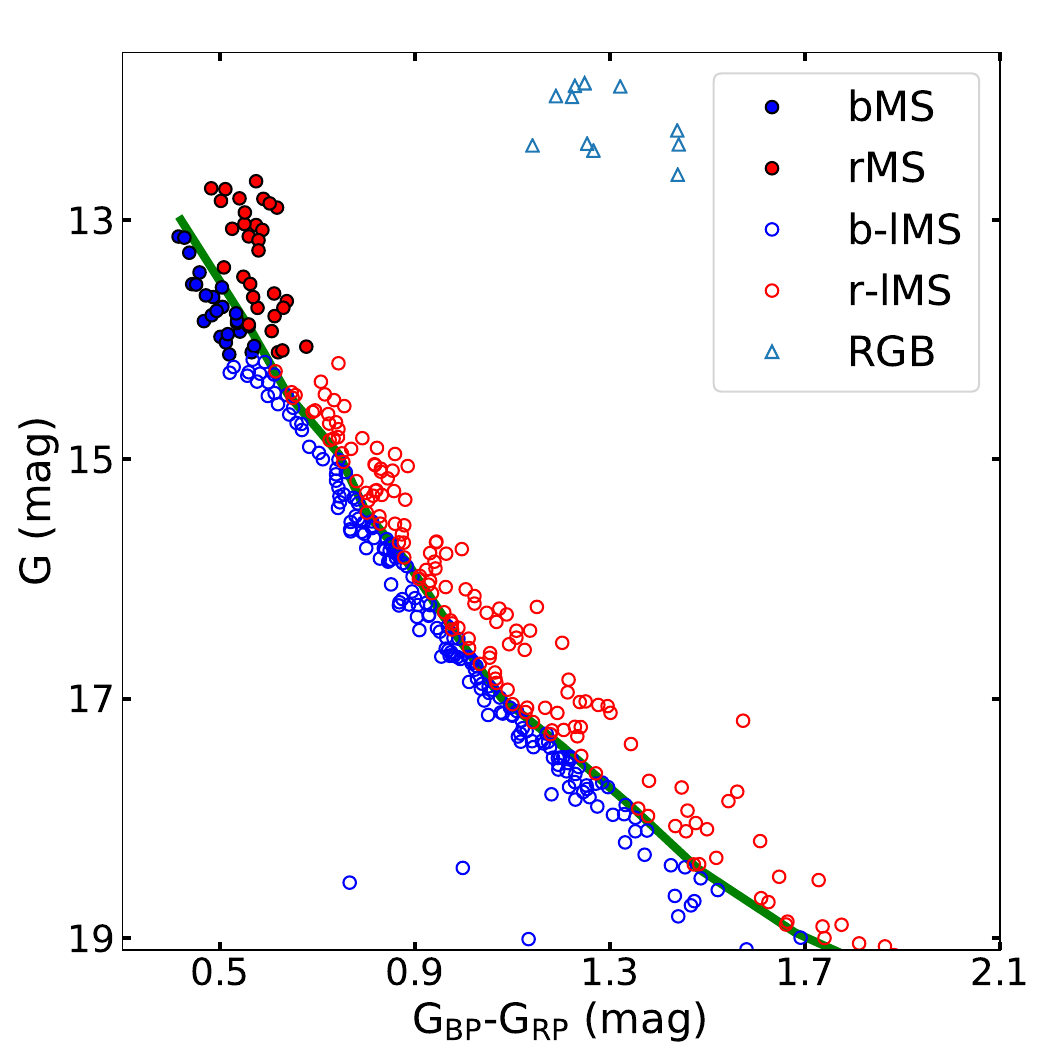}
    \caption{The colour-magnitude diagram with a green continuous line representing the fiducial line of main sequence stars. The bMS and rMS stars belonging to the extended Main Sequence Turn-Off region of NGC 2355 are shown by blue and red points, respectively. Blue and red open circles denote the b-lMS and the r-lMS stars of the lower main sequence population.}
    \label{fig:fiducial}
\end{figure}
For further investigation, we divided the population of eMSTO stars into two groups. The eMSTO stars below the fiducial line were grouped as bMS stars, whereas those above the fiducial line were called rMS stars. We divided the MS into magnitude bins of 0.5 mag to derive the fiducial line. The fiducial line is the interpolation of the median of magnitude and colour of these magnitude bins. The bMS and rMS stars within the eMSTO regions are shown in Figure~\ref{fig:fiducial}.  There were 23 bMS stars and 31 rMS stars in the eMSTO region. We grouped MS stars beyond the eMSTO region as lower main sequence (lMS) stars.  The red giant branch stars of the CMD are grouped as RGB stars. We divided the lMS stars into two groups: the lMS stars with colour bluer than the fiducial line as b-lMS stars and stars redder than the fiducial line as r-lMS stars. We have used these sub-samples of bMS, rMS, lMS, b-lMS, and r-lMS in the subsequent analysis in the paper. 

\subsection{Physical properties}\label{phy_param}
\subsubsection{Cluster properties through photometric analysis}
We used extinction, A$_{V}$, values for the cluster members given by  \citet{2022A&A...658A..91A} to calculate the reddening, E(B-V), values using relation E(B-V)=A$_{V}$/R$_{V}$ where R$_{V}$ is the total-to-selective extinction, taken to be 3.1, for the diffused interstellar medium \citep{1989ApJ...345..245C}. We calculated the mean extinction, A$_{V}$, value to be 0.346$\pm$0.123 mag which corresponds to E(B-V) = 0.112$\pm$0.040 for NGC 2355. \citet{2022A&A...658A..91A} used the broadband photometric data from \textit{Gaia} EDR3, Pan-STARRS1, SkyMapper,  AllWISE, and 2MASS to calculate the A$_{V}$ values of stars with a typical precision of  0.13 mag up to G=14 mag. The calculated mean A$_{V}$ value agrees with the mean A$_{V}$ value of 0.323$\pm$0.018 estimated by \citet{2021MNRAS.504..356D} using \textit{Gaia} DR2 optical G$_{\rm BP}$ and G$_{\rm BP}$ bands data. We also calculated the mean reddening values for bMS and rMS stars to be 0.099$\pm$0.027 and 0.098$\pm$0.043 mag, respectively. These values are in excellent agreement which indicates the absence of differential reddening for the eMSTO stars in NGC 2355.

The distance of cluster NGC 2355 was determined through the parallax inversion method. We calculated the mean $\varpi$ from the member stars to be 0.522 $\pm$ 0.046 mas, translating into a distance of 1853 $\pm$ 84 pc for NGC 2355.  This distance was calculated after compensating the mean $\varpi$ of NGC 2355 for a systematic global parallax offset of -0.017 mas for \textit{Gaia} DR3 parallaxes estimated by \cite{2021A&A...649A...4L}. The distance modulus calculated from the distance of this cluster was found to be 11.34 $\pm$ 0.10 mag. The age of this cluster was estimated by fitting extinction corrected \citet{2017ApJ...835...77M} isochrones\footnote{http://stev.oapd.inaf.it/cgi-bin/cmd}  of the metallicity Z = 0.0163 on the CMD for the obtained distance modulus and extinction. The age of the best-fit isochrone was found to be 1 Gyr. We determined the mass of the individual member stars through the isochrone fitting on the CMD of the cluster. The CMD of NCG 2355 with the best-fit isochrone is shown in Figure~\ref{age} and the masses of the individual member stars are given in Table~\ref{tab_param}. 

We used the multiple-part power law form of mass function provided by \citet{2001MNRAS.322..231K} to calculate the total mass of NGC 2355. We estimated the total mass of NGC 2355 as described by \citet{2009ApJ...700..506S}.  In this method, the total mass, M$_{tot}$ is calculated from the relation $M_{tot} = A\times \int_{m_{1}}^{m_{2}} M^{1-\alpha} dM$. The normalization constant, A, in this equation was estimated from the relation $N = A\times \int_{m_{1}}^{m_{2}} M^{-\alpha} dM$ for m$_{1}$ = 1.0 M$_{\odot}$, m$_{2}$ = 2.0 M$_{\odot}$, and N=218.  A brief description of this method can be found in our previous study \citet{2023arXiv230410138M}. Using this method, we obtained the total mass of the cluster to be 1.3$\pm$0.5 $\times$ 10$^{3}$ M$_{\odot}$ including members having masses above 0.08 M$_{\odot}$.

\subsubsection{Properties of individual stars through spectroscopic analysis}
\begin{figure}
	\includegraphics[width= 8.4 cm]{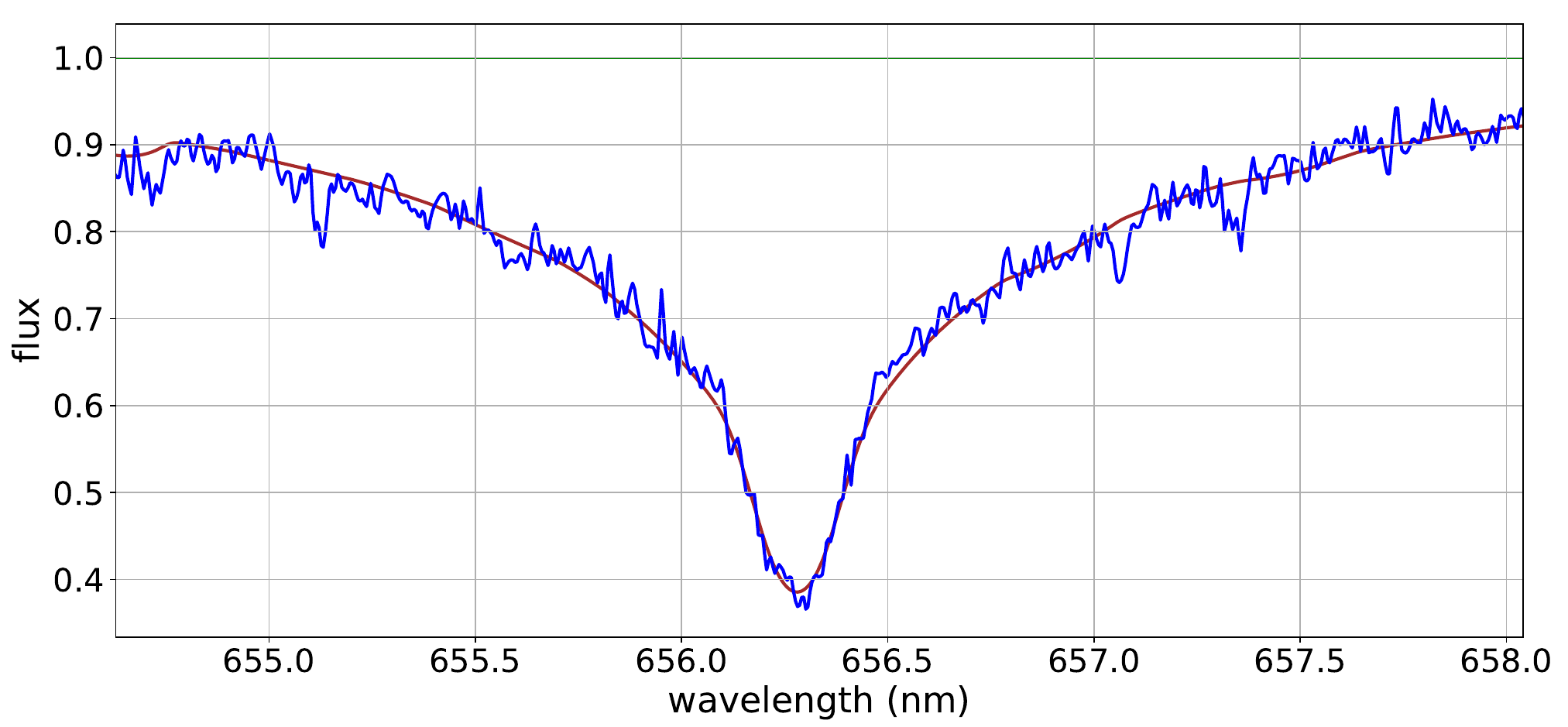}
        \includegraphics[width= 8.4 cm]{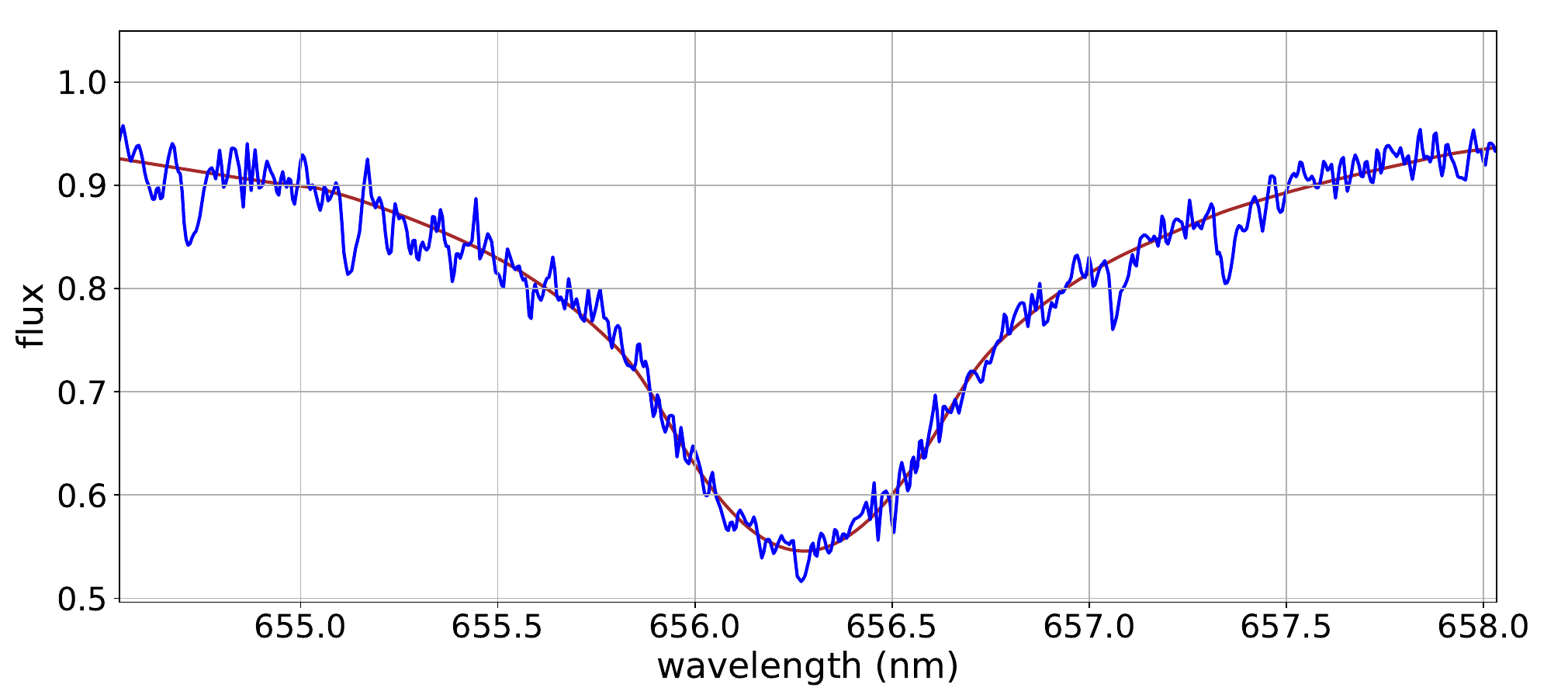}
    \caption{Plots showing synthetic spectra fitting on observed spectra. Blue curves denote the observed spectra, while the red curves exhibit the best-fit synthetic spectra. The upper panel corresponds to the slow-rotating (\textit{v} sin \textit{i} = 58.83$\pm$15.2 km s$^{-1}$) star with ID 41. The lower panel represents the best-fit plot for the fast-rotating (\textit{v} sin \textit{i} = 175.83$\pm$27.13 km s$^{-1}$) star with ID 28.}
    \label{fit_vsini}
\end{figure}

We used the \textit{iSpec} software solution package to estimate effective temperature (T$_{\rm eff}$) and projected rotational velocity (\textit{v} sin \textit{i}) of the stars \citep{2014A&A...569A.111B,2019MNRAS.486.2075B}. The \textit{iSpec} package provides various options for synthetic spectrum generation codes, line lists, and model atmospheres. We used the radiative transfer code SPECTRUM \citep{1994AJ....107..742G} and MARCS \citep{2008A&A...486..951G} atmosphere models available in the \textit{iSpec} to generate the synthetic spectra. Solar abundances for the synthetic spectra were taken from \citet{2005MSAIS...8...14K} models. We used original line lists provided in the SPECTRUM package covering 300 to 1100 nm. We fixed microturbulent velocity and the limb darkening to 2 km s$^{-1}$ and 0.6 for all the spectra, respectively. We obtained the best-fit synthetic spectra to the observed spectra through the global $\chi^{2}$ minimization of the parameters T$_{\rm eff}$, surface gravity log (g), and \textit{v} sin \textit{i} simultaneously. The value of a parameter corresponding to the best-fit synthetic spectra was considered as the estimated value of the parameter for the star. We have presented the illustrative plots of best-fit synthetic spectra on the observed spectra for slow and fast-rotating stars in Figure~\ref{fit_vsini}. The derived T$_{\rm eff}$ and \textit{v} sin \textit{i} values together with membership probabilities and other physical parameters of eMSTO stars in Table~\ref{tab_param}. The \textit{v} sin \textit{i} values we estimated for the eMSTO stars mostly agree with those reported by \citet{2022A&A...666A.121R}.

The high-resolution UVES spectra were available for RGB stars only. The \textit{v} sin \textit{i} and T$_{\rm eff}$ values for RGB stars were determined using these high-resolution spectra. The estimated \textit{v} sin \textit{i} values for RGB stars range from  1.23$\pm$0.45 to 7.98$\pm$0.21 km s$^{-1}$. We estimated \textit{v} sin \textit{i} and T$_{\rm eff}$ of eMSTO stars and lower MS stars from the medium-resolution GIRAFFE spectra. We found \textit{v} sin \textit{i} values for the eMSTO stars in the range 15.59$\pm$8.71 to 225.70$\pm$35.72 km s$^{-1}$. These spectroscopically estimated values of \textit{v} sin \textit{i} and T$_{\rm eff}$  are used in the following analysis of the present study.  

\begin{table*}
\caption{Physical parameters of the eMSTO stars in the cluster NGC 2355. The ID, right ascension, declination, G band magnitude,  (G$_{\rm BP}$-G$_{\rm RP}$) colour, membership probability, effective temperature, radial velocity, projected rotational velocity, mass, and the group of the eMSTO stars are given in columns 1, 2, 3, 4, 5, 6, 7, 8, 9, 10, and 11, respectively. The radial velocity of eMSTO stars given here is taken from \citet{2022A&A...666A.121R}.} \label{tab_param}
\begin{center}
\begin{tabular}{ccccccccccccccc}  
\hline
ID  &    RA (J2000)&   Dec. (J2000)&         G& (G$_{\rm BP}$-G$_{\rm RP}$)& prob.& T$_{\rm eff}$& RV & \textit{v} sin \textit{i}& Mass& Group \\
     & (deg)& (deg)& (mag)& (mag)& & (K)& (km s$^{-1}$)& (km s$^{-1}$)& (M$_{\odot}$)& \\
\hline     
  1& 109.08023& 13.95845& 13.976&  0.501& 0.99& -& -& -&  1.602& bMS\\
  2& 109.30089& 13.56054& 13.438& 0.458& 0.99& -& -& -& 1.77& bMS\\
  3& 109.27791& 13.55978& 13.138& 0.415& 0.99& -& -& -& 1.864& bMS\\
 4& 109.18357&  13.68092& 13.396&  0.508& 0.99& 7237.24$\pm$241.08& 37.7$\pm$1.4& 88.52$\pm$14.33& 1.784& rMS\\
  5& 109.26894& 13.61388& 13.042& 0.574& 0.72& 7051.75$\pm$321.58& 34.4$\pm$0.5& 109.58$\pm$18.45& 1.893& rMS\\
  6& 109.39146& 13.67027& 13.273& 0.437& 0.99& 7623.46$\pm$103.45& 35.5$\pm$2.2& 103.19$\pm$23.86& 1.822& bMS\\
  7& 109.26085& 13.65129& 13.845& 0.467& 0.96& 7325.04$\pm$105.44& 35.6$\pm$0.4& 29.32$\pm$10.96& 1.643& bMS\\
  8& 109.26866& 13.68081& 13.932& 0.541& 0.99& 7142.59$\pm$ 97.96& 36.9$\pm$0.4& 29.94$\pm$5.28& 1.615& bMS\\
  9& 109.2684& 13.69711& 13.536& 0.443& 0.99& 7686.74$\pm$ 76.27& 36.2$\pm$1.0& 66.24$\pm$14.29& 1.739& bMS\\
 10& 109.30153& 13.73161& 14.023& 0.512& 0.99& 7433.93$\pm$113.02& 35.1$\pm$2.8& 119.19$\pm$22.51& 1.59& bMS\\
 11& 109.2077& 13.73256& 13.928& 0.606& 0.99& 7314.85$\pm$144.88& 36.0$\pm$3.5& 120.65$\pm$23.27& 1.616& rMS\\
 12& 109.25139& 13.73758& 13.861& 0.535& 0.99& 7380.10$\pm$132.05& 34.2$\pm$2.0& 119.53$\pm$21.14& 1.637& bMS\\
 13& 109.2431& 13.7338& 13.803& 0.612& 0.95& 7131.14$\pm$77.65& 35.7$\pm$0.6& 37.64$\pm$7.95& 1.657& rMS\\
 14& 109.27174& 13.74128& 14.105& 0.565& 0.94& 7335.02$\pm$93.85& 30.8$\pm$4.7& 151.17$\pm$33.15& 1.562& bMS\\
 15& 109.2752& 13.74267& 12.742& 0.511& 0.99& 7308.84$\pm$ 86.71& 34.8$\pm$1.2& 103.25$\pm$16.23& 1.986& rMS\\
 16& 109.23403& 13.74478& 12.819& 0.54& 0.99& 7481.92$\pm$126.05& 37.8$\pm$0.3& 15.59$\pm$8.71& 1.963& rMS\\
 17& 109.25578& 13.75799& 13.474& 0.548& 0.99& 7413.67$\pm$117.7& 34.1$\pm$1.6& 103.43$\pm$17.67& 1.759& rMS\\
18& 109.24737& 13.75092& 12.823& 0.589& 0.98&  -& -& -& 1.962& rMS\\
 19& 109.2418& 13.76873& 12.676& 0.574& 0.99& 7223.79$\pm$115.39& 30.4$\pm$0.8& 74.18$\pm$11.70& 2.007& rMS\\
 20& 109.25306& 13.77097& 14.058& 0.677& 0.96& -& -& -& 1.576& rMS\\
21& 109.2206& 13.73247& 13.083& 0.587& 0.99& 7375.54$\pm$82.17& 37.5$\pm$2.5& 140.25$\pm$19.02& 1.882& rMS\\
 22& 109.28613& 13.75878&  12.84& 0.502& 0.95& 7237.07$\pm$103.48& 35.8$\pm$2.1& 162.36$\pm$22.91& 1.956& rMS\\
23& 109.29443&13.78722& 12.735& 0.482& 0.99& 7399.87$\pm$80.74& 35.5$\pm$1.2& 104.16$\pm$19.14& 1.988& rMS\\
 24& 109.26673& 13.77624& 13.137& 0.559& 0.99& 7199.52$\pm$127.5& 31.3$\pm$3.3& 183.79$\pm$27.57& 1.864& rMS\\
25& 109.26427& 13.79233& 13.073& 0.525& 0.99& -& -& -& 1.885& rMS\\
26& 109.28822& 13.79336& 13.614&  0.611& 0.98& 6976.67$\pm$86.18& 34.9$\pm$1.7& 140.87$\pm$26.19& 1.716& rMS\\
27& 109.29032& 13.8116& 13.563& 0.504& 0.99& -& -& -& 1.73& bMS\\
28& 109.33893& 13.81507& 13.726& 0.505& 0.99& 7230.92$\pm$158.99& 36.7$\pm$3.1& 175.83$\pm$27.13& 1.678& bMS\\
 29& 109.36126& 13.81018& 13.147& 0.427& 0.99& 7531.43$\pm$241.3& 35.0$\pm$0.6& 49.86$\pm$10.07& 1.861& bMS\\
 30& 109.31774& 13.80859& 13.887& 0.559& 0.99& 7022.35$\pm$175.82& 33.7$\pm$5.6& 216.26$\pm$32.93& 1.629& rMS\\
 31& 109.30771& 13.81213& 12.896& 0.617& 0.72& 6763.24$\pm$149.88& 35.9$\pm$2.8&  171.41$\pm$21.65& 1.938& rMS\\
32& 109.5021& 13.73835& 13.644& 0.486& 0.99& -& -& -& 1.706& bMS\\
 33& 109.44418& 13.79675& 13.033& 0.55&  0.99& 6956.8$\pm$102.86& 40.8$\pm$4.5& 218.78$\pm$29.29& 1.896& rMS\\
34& 109.4982& 13.80884& 14.123& 0.519& 0.99& -& -& -& 1.556& bMS\\
35& 109.20254& 13.76032& 13.678& 0.637& 0.90& 7073.72$\pm$359.0& 33.0$\pm$3.7& 159.66$\pm$28.71& 1.695& rMS\\
36& 109.21866& 13.76036& 13.168& 0.579& 0.65& 7324.16$\pm$104.82& 34.0$\pm$3.1& 178.06$\pm$25.46& 1.854& rMS\\
37& 109.2233& 13.77248& 12.861& 0.602& 0.99& 7471.63$\pm$117.22& 35.3$\pm$2.6& 142.33$\pm$25.53& 1.949& rMS\\
 38& 109.23429& 13.78643& 13.734& 0.577& 0.99&  7770.8$\pm$148.63& 50.2$\pm$0.3& 18.63$\pm$5.87& 1.676& rMS\\
39& 109.21833& 13.80952& 13.843& 0.535& 0.99& -& -& -& 1.643& bMS\\
40& 109.15049& 13.81682& 13.873& 0.559& 0.99& 7213.64$\pm$158.31& 36.9$\pm$0.9& 48.66$\pm$10.38& 1.633& rMS\\
41& 109.13963& 13.83815& 13.796& 0.483& 0.99& 7289.45$\pm$163.45& 36.0$\pm$1.0& 58.83$\pm$15.20& 1.659& bMS\\
42& 109.24741& 13.81411& 13.644& 0.568& 0.99&  -& -& -& 1.706& rMS\\
43& 109.26402&  13.82266& 14.053& 0.57& 0.99& 7425.65$\pm$88.5& 35.5$\pm$0.6& 39.43$\pm$14.58& 1.58& bMS\\
 44& 109.22606& 13.82122& 13.78& 0.533& 0.99& 7410.05$\pm$72.32& 34.8$\pm$1.3& 76.87$\pm$17.36& 1.662& bMS\\
 45& 109.23284& 13.83363& 14.105& 0.619& 0.98& 7112.6$\pm$115.56& 33.0$\pm$5.6& 187.02$\pm$31.85& 1.562& rMS\\
46& 109.31758& 13.85434&  13.539& 0.451& 0.99& 7662.58$\pm$71.91& 36.7$\pm$0.9& 50.60$\pm$15.25& 1.738& bMS\\
47& 109.31989& 13.86067& 13.733& 0.63& 0.99& -& -& -& 1.676& rMS\\
48& 109.17429& 13.85847& 13.63& 0.471& 0.99& -& -& -& 1.711& bMS\\
49& 109.28437& 13.90664& 14.09& 0.628& 0.99& 7282.1$\pm$189.4& 35.1$\pm$2.2& 96.13$\pm$20.56& 1.567& rMS\\
50& 109.21547& 13.906& 12.937& 0.551& 0.99& 7018.4$\pm$119.73& 33.0$\pm$2.5& 163.17$\pm$23.53& 1.927& rMS\\
51& 109.11116& 13.85945& 13.954& 0.516& 0.99& -& -& -& 1.608& bMS\\
52& 109.03448& 13.89163& 13.759& 0.493& 0.99& -& -& -& 1.668& bMS\\
53& 109.20019& 13.92274& 13.534&  0.562& 0.98& 7120.89$\pm$166.06& 40.6$\pm$6.0& 225.70$\pm$35.72& 1.739& rMS\\
54& 109.23248& 13.94563& 13.254& 0.579& 0.99& 7041.03$\pm$133.55& 37.5$\pm$1.1& 180.02$\pm$24.30& 1.828& rMS\\
\hline

\end{tabular}
\end{center}
\end{table*}

\subsection{Unresolved binaries}

\begin{figure}
 \includegraphics[width= 8.4 cm]{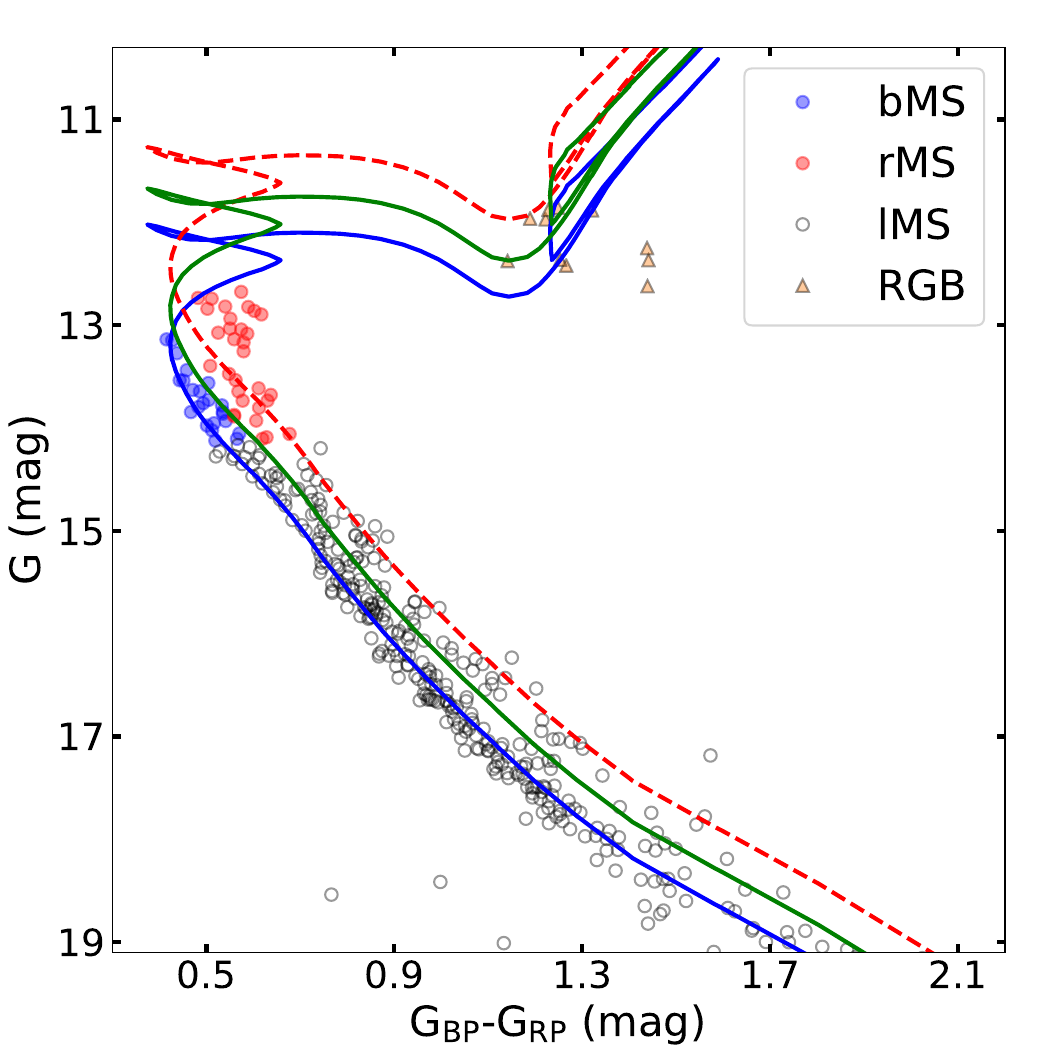}
    \caption{Colour-magnitude diagram of NGC 2355 fitted with \citet{2017ApJ...835...77M} single stars isochrone corresponding to 1 Gyr age shown by the blue continuous curve. The dashed red curve represents the same isochrone shifted by 0.752 mag in G bands to fit the equal mass binary sequence of the cluster NGC 2355. The green continuous curve represents the binary sequence for the 0.8 mass ratio binaries.}
    \label{fig:cmd_binary_seq}
\end{figure}

The presence of the binary stars can also influence the morphology of the CMD of the clusters as an unresolved binary would appear redder and brighter than the single star of a mass similar to the mass of the primary component on the MS. This redder and brighter shift of binaries may resemble the eMSTO in the upper MS of the clusters. The magnitude of the binary system can be expressed as follows:
\begin{center}
$m_{\rm binary} = m_{\rm p}$ - 2.5 log $\Bigg(1 + \dfrac{F_{\rm s}}{F_{\rm p}}\Bigg)$ 
\end{center}
where $m_\textrm{binary}$ and $m_\textrm{p}$ are the magnitudes of the whole binary system and the primary star, respectively. $F_\textrm{p}$ and $F_\textrm{s}$ denote the flux of the primary and the secondary stars of the binary system, respectively. The shift will be largest for the equal mass binaries as these binaries will be -2.5$\times$log(2) $\sim$ 0.752 mag brighter than the single star on the MS.  Therefore, we investigated the CMD of the cluster NGC 2355 by over-plotting \citet{2017ApJ...835...77M} isochrone of the metallicity Z = 0.0163 corresponding to the equal mass binaries as shown in Figure~\ref{fig:cmd_binary_seq}. The stars lying below G = 16.0 mag on the MS, especially between the G = 16.0 and G = 17.5 mag, form a narrow MS separated from the binary sequence. The gap between the single stars sequence and equal mass binaries sequence becomes narrower towards the upper MS, especially near the turn-off region. The two sequences mostly contain the distribution of the stars on MS of the CMD, however, the majority of the rMS stars in the upper eMSTO region are intriguingly beyond the equal mass binary sequence. This indicates that the apparent colour shift due to unresolved companion will not be able to produce most of the spread in the red part of the eMSTO in the CMD of NGC 2355. However, the low mass ratio binaries can still be responsible for the spin-down of bMS stars of the eMSTO \citep{2015MNRAS.453.2637D}. The bMS stars in the CMD are well contained within the single star sequence and binary sequence corresponding to mass ratio q=0.8 as shown in Figure~\ref{fig:cmd_binary_seq}. The magnitude shift for the binary sequence of mass ratio q = 0.8 was calculated to be 0.350 mag by using the luminosity-mass relation provided by \citet{2018MNRAS.479.5491E} in the above equation for the magnitude of the binary system. We have discussed the spin-down scenario of the fast-rotating stars due to tidal-locking in the low mass ratio binaries present in the bMS population in Section~\ref{tidal_lock}. 

\section{Origin of the \texorpdfstring{\MakeLowercase{e}}{e}MSTO in the cluster}

\subsection{Spread in the rotation rates}
\begin{figure*}
	\includegraphics[width= 17 cm]{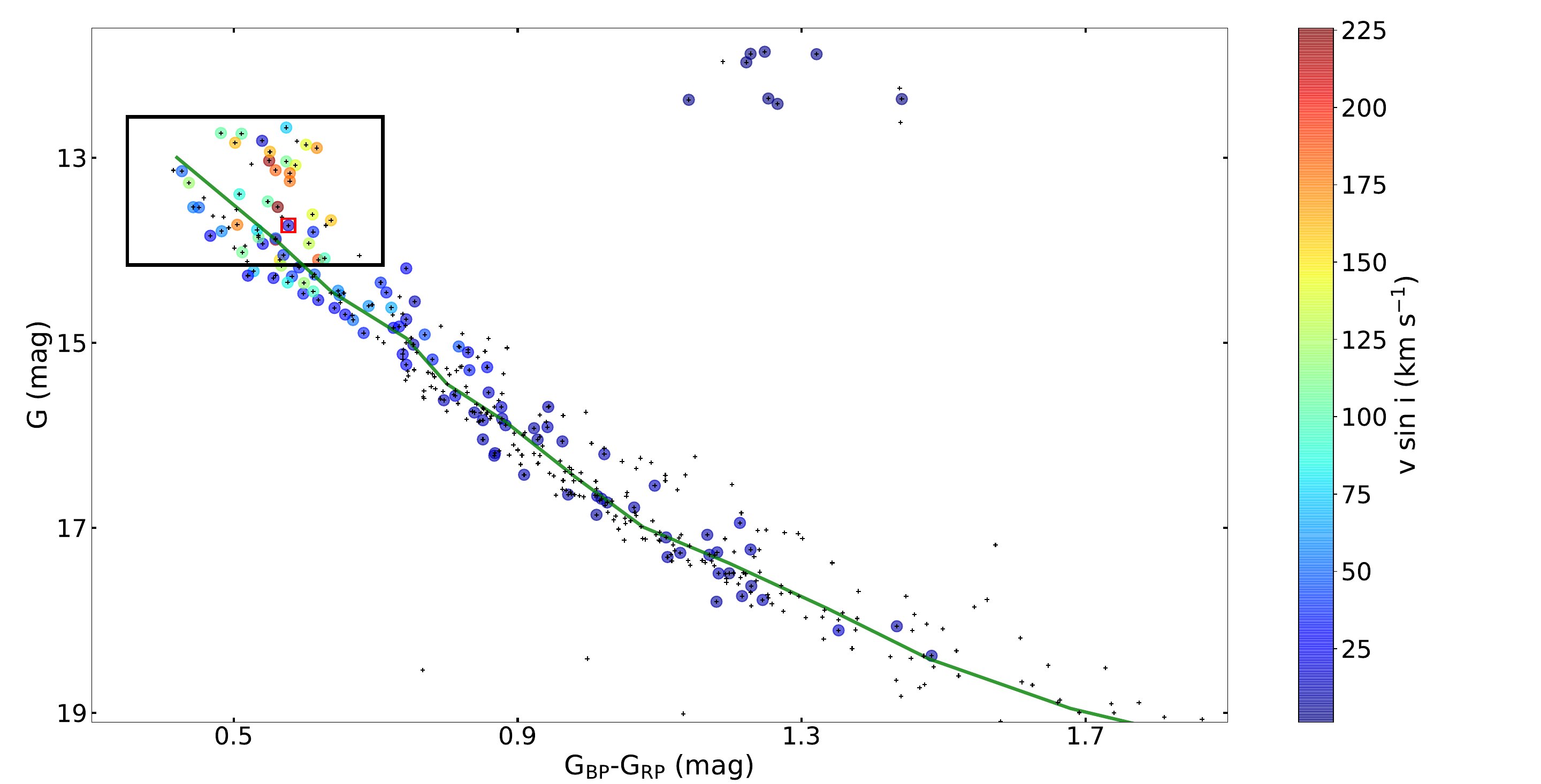}
    \caption{Plot of the colour-magnitude diagram colour-coded by projected rotational velocities \textit{v} sin \textit{i} of stars in NGC 2355. The green continuous line represents the fiducial line for main sequence stars. The rMS star with significantly high radial velocity (ID 38) is marked by a red square.}
    \label{fig:v_sini}
\end{figure*}
\begin{figure}
	\includegraphics[width= 8.3 cm]{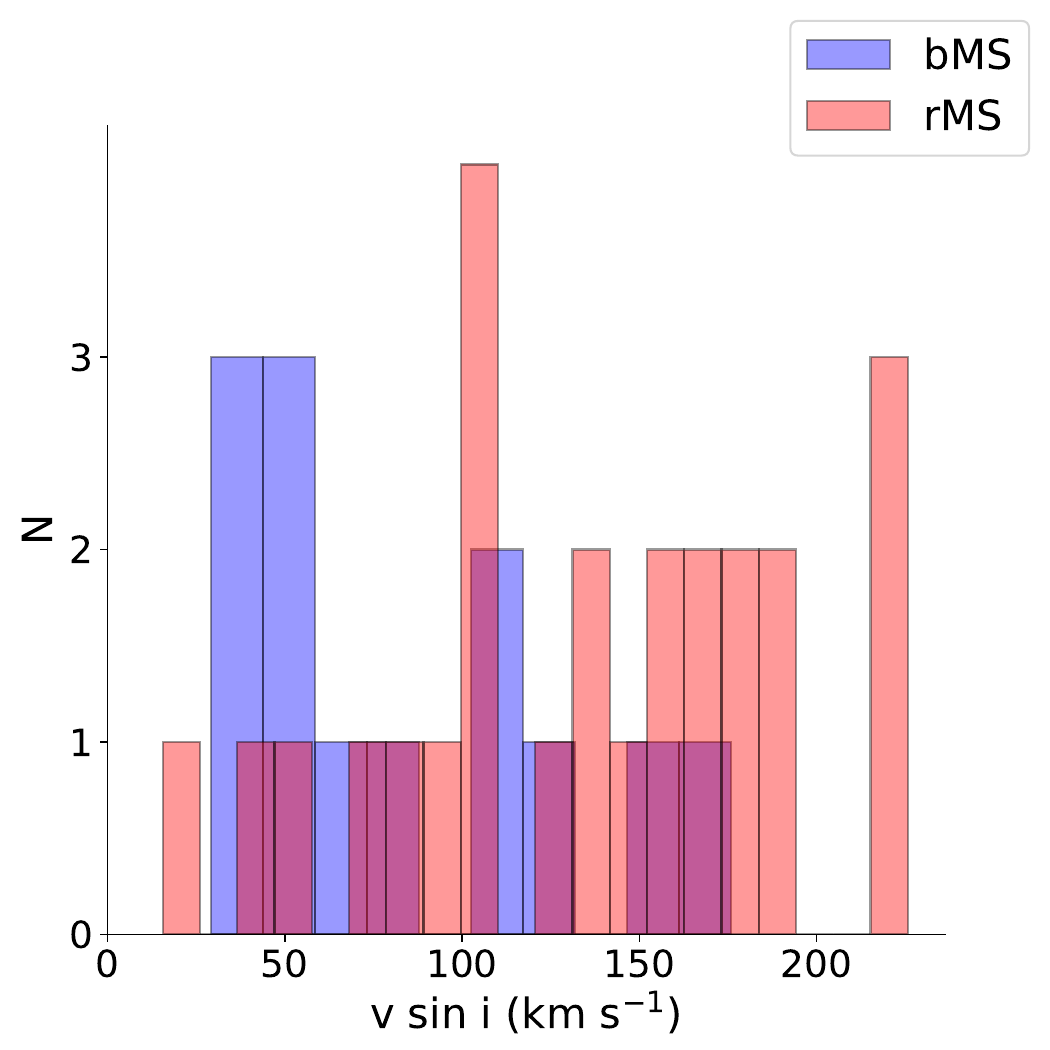}
    \caption{Histogram of projected rotational velocity, \textit{v} sin \textit{i}, distribution for bMS and rMS stars of the eMSTO in NGC 2355 by the blue and the red bars, respectively.}
    \label{fig:hist_vsini}
\end{figure}

The spread in the rotation rates of stars has also been suggested to be a possible reason behind the origin of eMSTO in the Galactic open clusters \citep{2018MNRAS.480.3739B,2019NatAs...3...76L}. We have shown the \textit{v} sin \textit{i} distribution of the stars on the CMD of the cluster NGC 2355 in Figure~\ref{fig:v_sini}. We found no clear-cut distinction between the fast and slow-rotating stars on the lower MS of NGC 2355 as these stars are mostly slow rotators. However, there is a conspicuous preferential distribution of fast-rotating stars on the red part and slow-rotating stars on the blue part of the CMD in the eMSTO region of the cluster. We found the mean \textit{v} sin \textit{i} values for the bMS and rMS stars of the eMSTO to be 81.3$\pm$5.6 and 135.3$\pm$4.6  km s$^{-1}$, respectively. We have shown the histogram of the \textit{v} sin \textit{i} values for the bMS and rMS stars in Figure~\ref{fig:hist_vsini}. It can be noticed from the histogram that the bMS stars have projected rotational velocities up to $\sim$176 km s$^{-1}$ whereas the rMS stars possess \textit{v} sin \textit{i} values up to $\sim$226 km s$^{-1}$.  Thus, the rMS stars exhibit a much wider range of rotational velocities, as reported in the previous study on the other clusters hosting eMSTOs \citep{2018AJ....156..116M}. However, the mean \textit{v} sin \textit{i} value for the stars belonging to the lower MS stars was found to be 26.5$\pm$1.3  km s$^{-1}$.
\begin{figure}
	\includegraphics[width= 8.3 cm]{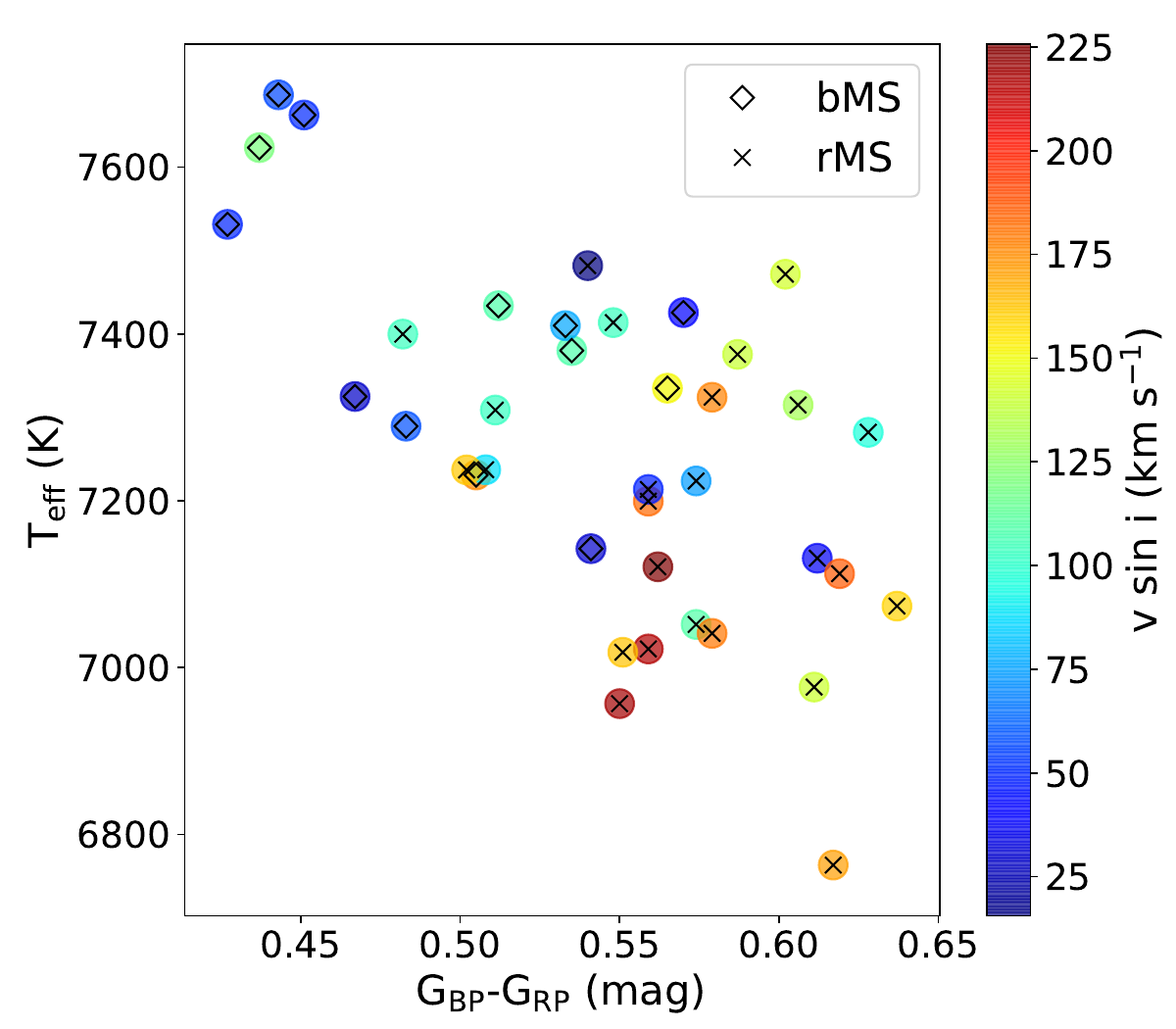}
    \caption{Plot of the effective temperature distribution as a function of the (G$_{\rm BP}$ - G$_{\rm RP}$) colour for the eMSTO stars colour-coded by their projected rotational velocities.}
    \label{fig:gbr_teff}
\end{figure}
The effects of stellar rotation, such as gravity darkening and dust-like extinction from the circumstellar excretion disc in the fast-rotating stars, may lead to a spread in the colour of the eMSTO stars on the CMD. The gravity-darkening effect can cause a decrease in the T$_{\rm eff}$ values for the fast-rotating stars. We studied the distribution of effective temperature, T$_{\rm eff}$, as a function of the G$_{\rm BP}$ - G$_{\rm RP}$ colour for eMSTO stars in NGC 2355. We found average  T$_{\rm eff}$ values of 7421.3$\pm$34.8 and 7190.1$\pm$32.0 K for the bMS and rMS stars, respectively. We also investigated the relation between stellar rotation and the effective temperature of the eMSTO stars in the cluster. We have shown the projected rotational velocity in a T$_{\rm eff}$-colour diagram in Figure~\ref{fig:gbr_teff}. We found that the fast-rotating rMS stars generally tend to have lower T$_{\rm eff}$ values than their slow-rotating counterparts. The gravity-darkening also includes the inclination effect which causes the fast-rotating stars to appear bluer and brighter for the pole-on view compared to the equator-on view \citep{1924MNRAS..84..665V,2020MNRAS.492.2177K}. Thus, the combined effect of the viewing angle and gravity darkening might contribute to the emergence of the eMSTO in cluster NGC 2355.

\subsection{Dust-like extinction from the circumstellar excretion disc}
\begin{figure*}
    \includegraphics[width= 17 cm]{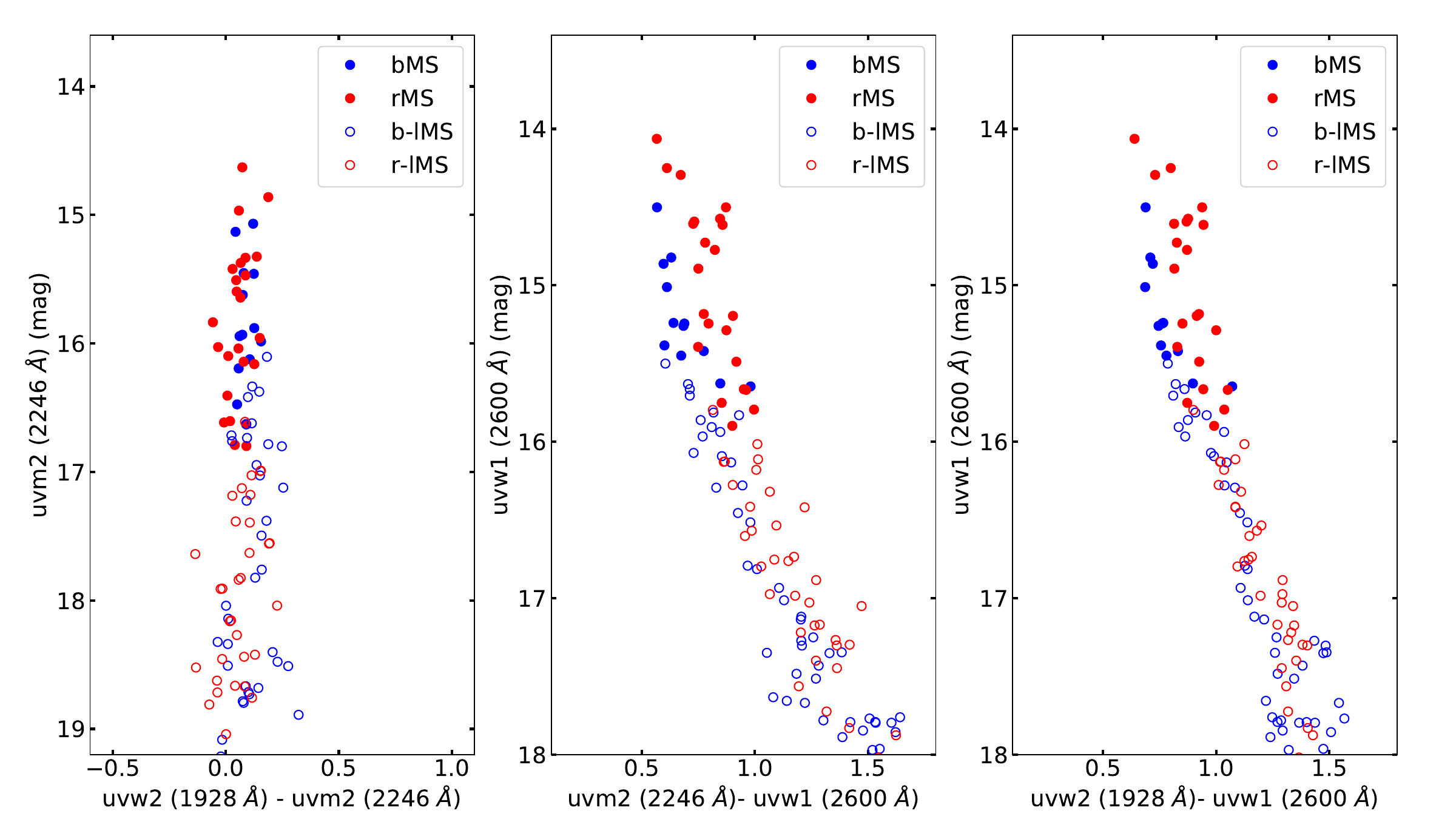}
    \caption{The colour-magnitude diagrams of NGC 2355 using near ultraviolet uvw2, uvm2, and uvw1 bands magnitudes from the \textit{Swift} survey archive. The left panel is the uvm2 versus (uvw2 - uvm2) diagram; the middle panel is the uvw1 versus (uvm2 - uvw1) diagram; and the right panel is the uvw1 versus (uvw2 - uvw1) diagram for stars in NGC 2355.}
    \label{fig:cmd_UV}
\end{figure*}
The material ejected from the fast-rotating stars forms the circumstellar excretion disc. The dust-like extinction from the excretion disc of the stars can cause fast-rotating stars to appear redder \citep{2023MNRAS.521.4462D}. The absorption from the disc combined with the viewing angle may also produce a broadening of the upper MS, resulting in the eMSTO. This dust-like extinction is expected to be wavelength-dependent and can be detected in the ultraviolet CMD \citep{2023A&A...672A.161M,2023MNRAS.521.4462D}. We used the uvw2 (1928 \AA), uvm2 (2246 \AA), and uvw1 (2600 \AA) bands ultraviolet data to investigate the dust-like extinction properties of the eMSTO stars in the cluster NGC 2355. The ultraviolet CMDs for NGC 2355 are shown in Figure~\ref{fig:cmd_UV}. We have shown the stars in the blue and the red parts of the eMSTO and the lower main sequence, lMS, by different groups in the figure to compare the UV extinction from them. The broadened upper MS is conspicuous in the UV CMDs except for uvm2 versus (uvw2-uvm2) CMD. We noticed an interesting case where stars belonging to the red part of the MS, including both the rMS and the r-lMS stars, of the optical CMD are bluer than their counterparts in the uvm2 versus (uvw2-uvm2) UV CMD. We did not find substantial evidence of any excess UV extinction from rMS stars in the eMSTO sample compared to the r-lMS stars in lower MS in NGC 2355 to support the hypothesis of dust-like extinction from the fast-rotating stars suggested by \citet{2023MNRAS.521.4462D}.

\subsection{Possible mechanisms for spread in rotation rates}
\subsubsection{Tidally-locked binaries}\label{tidal_lock}
The low mass ratio unresolved binaries may be present in the bMS population of the eMSTO which may have slowed down due to tidal locking. The tidal locking synchronizes the rotation rates of the primary and secondary stars in a binary system with their orbital motion \citep{2013ApJ...764..166D}. The synchronization may slow down the stellar rotation by effectively transferring the angular momentum from the rotational motion to the orbital motion of the system. The effect of the synchronization of the rotation of the stars can be investigated by calculating the required synchronization time. We estimated the synchronization time, $\tau_{sync}$, by following \citet{2002MNRAS.329..897H} formulae for MS binary stars with radiative envelopes having primary star with mass $\geq$ 1.25 M$_{\odot}$. The formula can be expressed as follows:
\begin{center}
$\dfrac{1}{\tau_{sync}}=5\times 2^{5 / 3}\left(\dfrac{G M_{\rm p}}{R_{\rm p}^3}\right)^{1 / 2} \dfrac{M_{\rm p} R_{\rm p}^2}{{I_{\rm p}}} q^2\left(1+q\right)^{5 / 6} E_2\left(\dfrac{R_{\rm p}}{a}\right)^{17 / 2}$
\end{center}
where $M_\textrm{p}$, $R_\textrm{p}$, and $I_\textrm{p}$ are the mass, radius, and moment of inertia of the primary stars, respectively. The $G$ in the equation denotes the gravitational constant. $q$ represents the mass ratio of binary systems, while $a$ denotes the separation between binary components. The physical quantities in the above equation are in the CGS units system. $E_\textrm{2}$ is the second-order tidal coefficient. The value of $E_\textrm{2}$ can be calculated using the following relation provided by \citet{2019ApJ...876..113S} derived through fitting the values of $E_\textrm{2}$ and mass of stars given in the original study by \citet{1975A&A....41..329Z}:
\begin{center}
   $ E_{\rm 2} = 1.592 \times 10^{-9} M_{\rm p}^{2.84} $
\end{center}
 where $M_\textrm{p}$ is in the solar unit M$_{\odot}$. We estimated the radius of the primary star by using the empirical relations $R_\textrm{p}$  $\approx$ 1.06 M$^{0.945}$ for $M_\textrm{p}$ < 1.66 M$_{\rm \odot}$ and $R_\textrm{p}$  $\approx$ 1.33 M$^{0.555}$  in case of $M_\textrm{p}$ > 1.66 M$_{\rm \odot}$ provided by \citet{1991Ap&SS.181..313D}.
\begin{figure*}
	\includegraphics[width= 17 cm]{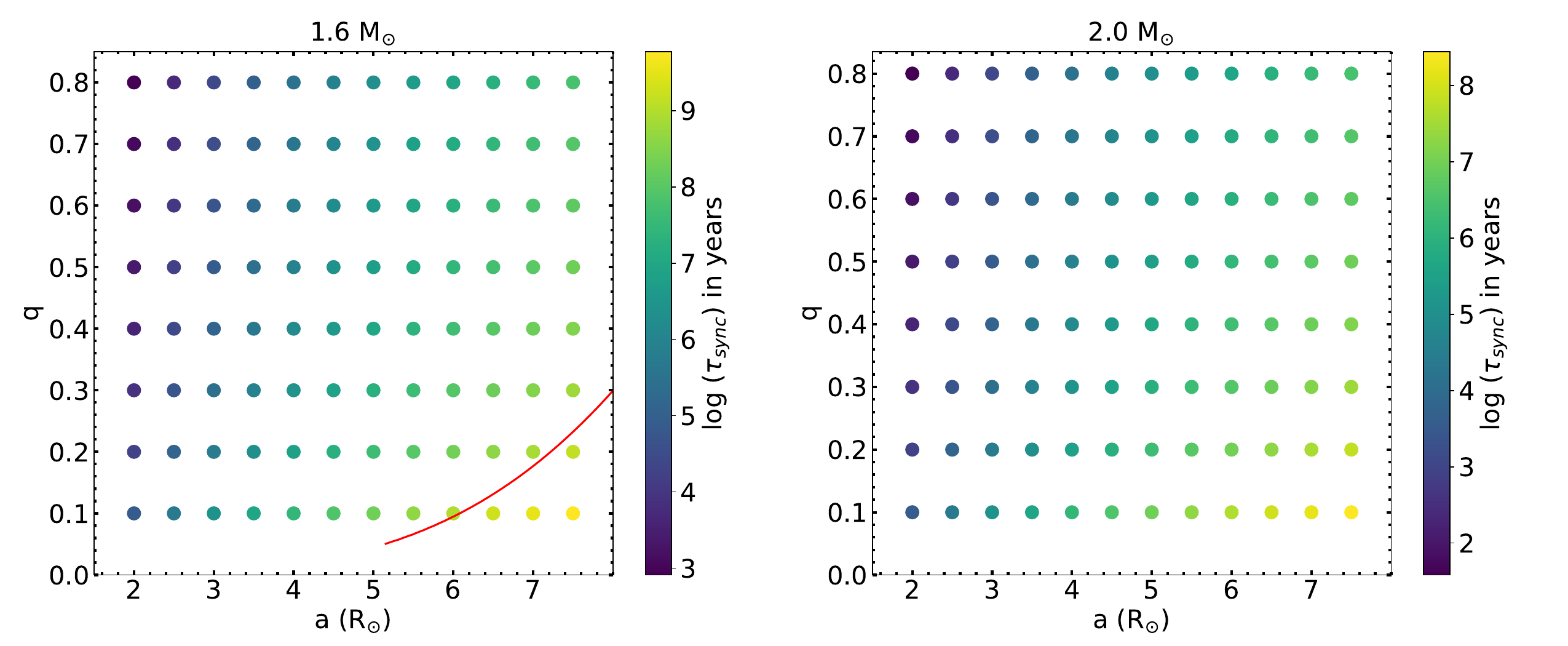}
    \caption{The plot showing synchronization time ($\tau_{sync}$) distribution among bMS stars if they were binaries of different mass ratios and separations. The plot exhibiting $\tau_{sync}$ distribution for a typical binary with the primary star mass of 1.6 M$_{\rm \odot}$ is shown in the left panel while the right panel plot illustrates $\tau_{sync}$ values for a 2.0 M$_{\rm \odot}$ primary star binary system. The points are colour-coded for their corresponding synchronization time values. The red continuous curve, in the left panel, marks the binary components separation boundary for $\tau_{sync}$ = 1000 Myr, same as the cluster NGC 2355 age.}
    \label{sync_time}
\end{figure*}
We have shown the impact of the mass ratio and separation between binary components on the synchronization time of the binary systems in Figure~\ref{sync_time}. From the above relations, we can deduce that the binary system with a higher primary star mass would have a shorter synchronization time than the binary system with a relatively lower primary star mass. Since the eMSTO stars in our sample have masses from $\sim$1.6 to 2.0 M$_{\rm \odot}$ so, we have shown plots of the synchronization time distribution only for the two bounding masses, i.e., 1.6 M$_{\rm \odot}$ and 2.0 M$_{\rm \odot}$ in the figure. We noticed that the $\tau_{sync}$ sharply increases with the separation between the binary components for any particular mass ratio. The binary system with a larger separation between the components would have weaker tidal torque and hence, longer synchronization time. We also found that the $\tau_{sync}$ values for low mass ratio binaries were greater than the $\tau_{sync}$ for the relatively high mass ratio binaries of the same mass and separation. The synchronization time for the close binary systems (a < 7 R$_{\rm \odot}$) with q $\geq$ 0.2 is less than the age of the cluster NGC 2355. For binaries with q < 0.2 also, the $\tau_{sync}$ is generally less compared to the age of the cluster up to a separation of $\sim$6 R$_{\rm \odot}$. We expect that most of the close binary systems in the bMS population of the eMSTO stars may have synchronized rotational and orbital motions due to tidal locking and thus become slow-rotating as suggested by \citet{2017NatAs...1E.186D}. 
\begin{figure}
	\includegraphics[width= 8 cm]{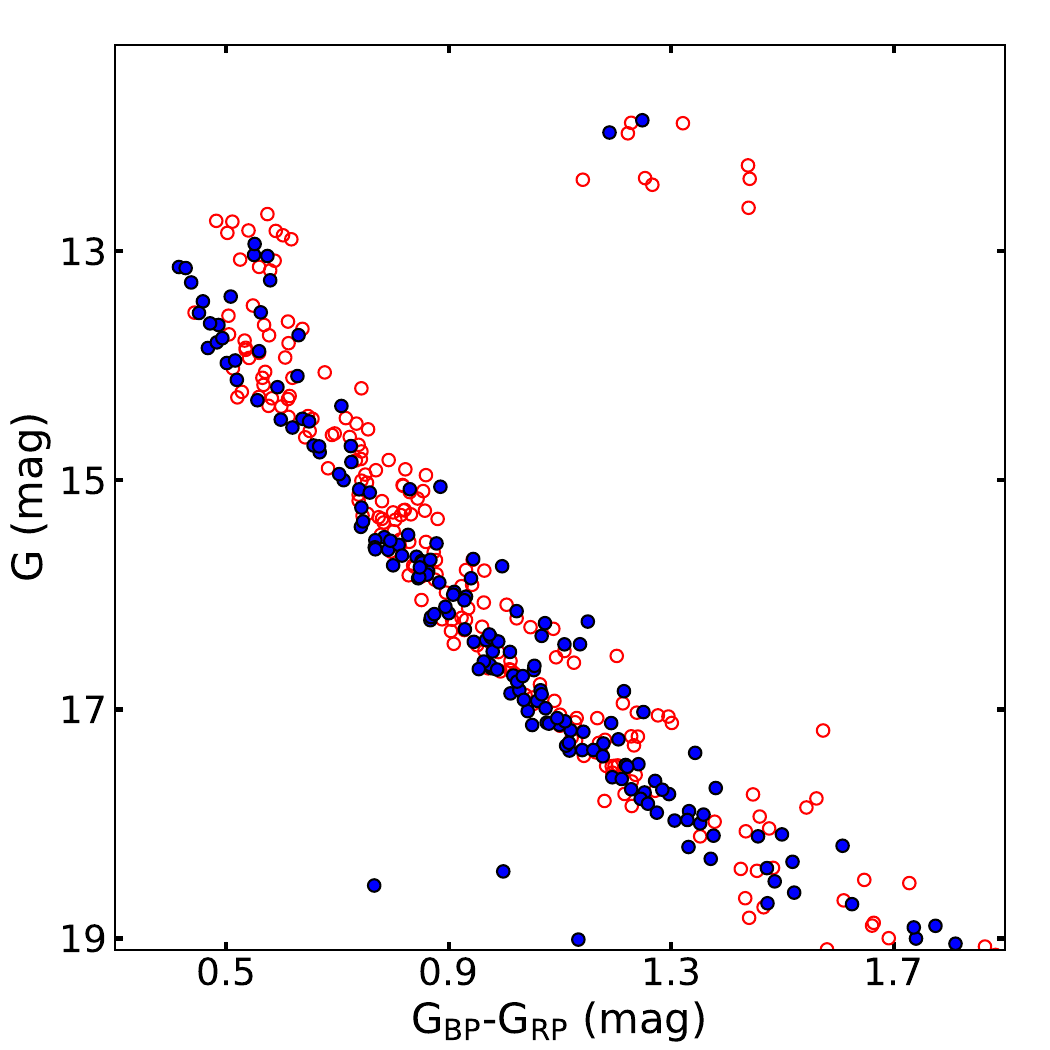}
    \caption{The colour-magnitude diagram of stars in the NGC 2355 cluster represents inner and outer region stars with red open circles and blue filled circles, respectively.}
    \label{rad_distri}
\end{figure}

The radial distribution of the eMSTO stars in NGC 2355 is crucial for assessing the possibility of populating blue eMSTO stars from tidally braked slow-rotating stars. The tidally braked slow-rotating low mass ratio binaries of the bMS population are expected to be preferentially concentrated in the central region due to the mass segregation of the binary systems. To investigate the radial distribution of the bMS and rMS stars, we divided the stars of NGC 2355 into two groups: inner region and outer region stars. The circular region around the cluster center with a radius equal to the half mass radius of 0.104$\degree$ provided by \citet{2022AJ....164...54Z} was taken as the inner region. The remaining annular region was considered the outer region of the cluster. The inner and outer regions had 224 and 187 stars, respectively. We have shown the locations of these two different populations in the CMD of NGC 2355 in Figure~\ref{rad_distri}. We found that the bMS stars of the eMSTO were preferentially located in the outer region of the cluster. This contradicts our expectation that bMS stars should be preferentially segregated in the inner region if they were tidally locked binary systems. Therefore, bMS stars seem less likely to be the tidally braked low mass ratio binaries. Similarly, a few previous studies also found that the stars belonging to the red part of the eMSTO were preferentially inner region stars \citep{2021AJ....162...64M,2023MNRAS.518.1505K}.  As discussed in the previous section, only the low mass ratio binaries with massive primary stars can undergo a rapid tidal locking process. Based on findings from simulated populations, \citet{2023ApJ...949...53W} suggested that low mass ratio binaries will have primary stars heavier than 5 M$_{\rm \odot}$ for a population of 100 Myr age. In the case of clusters older than 100 Myr, the primary star in the low mass ratio binary would have already evolved off the MS \citep{2023ApJ...949...53W}. Thus, the cluster NGC 2355 (age $\sim$ 1000 Myr) may not possess a tidally locked low mass ratio binary in its bMS population.

\subsubsection{Star-disc interaction}\label{sec:rad_distri}
\citet{2020MNRAS.495.1978B} suggested that the spread in rotation may be caused by differences in the star-disc interaction during the PMS phase. The coupling between the star and the protoplanetary disc is thought to be responsible for extracting a large amount of angular momentum from the star \citep{1993AJ....106..372E,2023A&A...678A...7A}. However, as soon as the disc is dissipated or decoupled from the star, the latter is free to spin up as it contracts to the MS. Therefore, an early spin-up leads to a faster rotation on the MS since the star can spin up for a longer time. Many factors can affect the duration of the star-disc coupling. The disc can be dispersed by its accretion onto the star, dynamical encounter, photo-evaporation from the star itself, or surrounding massive stars \citep{2021MNRAS.508.3710R}. The distribution of the rotation rates of these stars thus strongly depends on the environment during the formation stage. If the stars in a cluster face a higher rate of disc destruction, the cluster will host a larger number of fast-rotating stars leading to a change in its main sequence turn-off morphology.

The stars in the central part of the cluster may face higher destruction of their disc due to dynamical interactions caused by higher stellar density and the photoionization from the larger fraction of massive stars towards the center \citep{2024AJ....167..120V}. The greater fraction of the destruction of the protoplanetary disc of stars may result in a larger fraction of the fast-rotating stars in the inner region of the cluster \citep{2020MNRAS.495.1978B}. It can also be inferred from Figure~\ref{rad_distri} that the rMS stars of the eMSTO, which comprise mostly fast-rotating stars, are preferentially located in the inner region of the cluster NGC 2355. Thus, our findings of the radial distribution of eMSTO stars in NGC 2355 indicate that the star-disc interaction is most likely responsible for the observed spread in rotation rates of the eMSTO stars.

\subsection{The lower mass limit for fast rotation}

 We found that fast-rotating stars were brighter than G$\sim$14.16 mag in NGC 2355 which corresponds to a mass of $\sim$1.56 M$_{\odot}$ for the metallicity Z = 0.0163 (see Figure~\ref{fig:v_sini}). This mass limit coincides with the mass range of $\sim$1.5-1.6 M$_{\odot}$ where magnetic braking starts slowing down the stars \citep{2018ApJ...864L...3G,2019A&A...622A..66G,2020MNRAS.492.2177K,2020MNRAS.495.1978B}. The stars with masses above this mass limit on the MS are expected not to be efficiently spun down as they possess very thin convective envelopes. This mass limit increases with the increase in the metallicity of clusters \citep{2019A&A...622A..66G}. The stars in the lower main sequence (below $\sim$1.56 M$_{\odot}$) of NGC 2355 were found to be rotating slowly with a mean projected rotational velocity of the 26.5$\pm$1.3 km s$^{-1}$. The lower MS is marked by a complete absence of fast-rotating stars indicating the magnetic braking of stellar rotation for the lMS stars in NGC 2355. 

The magnetic braking in low-mass stars is related to the presence of a thick convective envelope, which allows for the generation of a large-scale magnetic field through a dynamo process. The coupling of this large-scale field with the mass loss causes the stars to efficiently lose angular momentum and their spin rate to brake as they evolve. This likely explains both the absence of fast-rotating stars in the lower part of the CMD and the lesser spread of the MS as the structural effects of rotation become negligible at low rotation rates. The dynamo becomes inefficient above $\sim$1.5M$_{\odot}$ and the stars keep the fast rotation throughout their MS lifetime. The transition between the two regimes is known as the Kraft break \citep{1967ApJ...150..551K}.

\section{Summary}

We study the origin of the extended main sequence turn-off present in the Galactic open cluster NGC 2355. It is a low mass (1.3$\pm$0.5 $\times$ 10$^{3}$ M$_{\odot}$) cluster with age of $\sim$1 Gyr. The MS stars, except rMS stars in NGC 2355, are contained between the single star and the equal-mass binary sequences. The majority of the rMS stars in the eMSTO region are distributed well beyond the equal mass binary sequence, which discards the possibility that the unresolved binaries could resemble eMSTO in NGC 2355 (see Figure~\ref{fig:cmd_binary_seq}). We further investigate the projected rotational velocity distribution of the stars in the cluster, which reveals that fast-rotating stars preferentially populate the red part of the eMSTO. The \textit{v} sin \textit{i} values for bMS, rMS, and lMS stars are found to be 81.3$\pm$5.6, 135.3$\pm$4.6, and 26.5$\pm$1.3 km s$^{-1}$, respectively. The spread in rotation rates of stars may lead to the origin of eMSTO due to various effects related to stellar rotation, such as dust-like extinction from excretion disc and gravity-darkening. We examine the dust-like extinction scenario through ultraviolet CMD created from the \textit{Swift} near-ultraviolet data. We do not find any substantial evidence of the excess ultraviolet absorption in the fast-rotating rMS population of NGC 2355 to support the hypothesis that dust-like extinction from the circumstellar disc makes them appear redder on the MS compared to their slow-rotating counterparts. However, a careful inspection of the effective temperature of the stars hints toward the contribution of gravity darkening in the colour spread of the upper MS of NGC 2355. The spread in rotation rates of the eMSTO stars can be explained by mechanisms such as tidal interaction in the binary stars and star-disc interaction in the PMS phase of stars. The synchronization time for the likely low mass ratio close binaries belonging to bMS stars of NGC 2355 is mostly shorter enough for them to become slow-rotating stars through the tidal-locking process. To further inspect the tidal locking scenario, we analyze the radial distribution of the stars in NGC 2355. Against the general expectation that bMS stars should be preferentially located in the inner region if they were close binaries, we find that the bMS stars are mostly located in the outer region of the cluster. The possible cause for this discrepancy could be the absence of the low mass ratio close binaries in the bMS population. So, the tidal locking in the close binaries appears to be the less likely reason for the spread in rotation rates of the eMSTO stars in the NGC 2355 open cluster. The different star-disc interaction time in the PMS phase of the stars may also lead to the spread in rotation rates of the eMSTO stars. The longer the star-disc interaction time, the slower the rotation rate of the star would be. The early destruction of the protoplanetary disc of stars in the central region would be higher due to dynamical interactions and photoionization, which may lead to a higher concentration of the fast-rotating stars towards the center of a cluster. We also find that the rMS stars, mostly fast-rotating, are preferentially concentrated in the inner region of the cluster NGC 2355. Therefore, the star-disc interaction during the PMS phase seems to be the most likely mechanism for the spread in the rotation rates of the upper MS stars and, thus, the origin of the eMSTO in the open cluster NGC 2355. We also notice an absence of fast-rotating stars in the lower main sequence beyond $\sim$1.56 M$_{\rm \odot}$ possibly because of the magnetic braking that effectively spins down their rotations. Further radial velocity analysis of the eMSTO stars involving high-resolution multi-epoch data will help us better understand the role of binary stars in the origin of the eMSTO in NGC 2355 and open clusters in general.

\section*{Acknowledgements} 

This research was supported by the Chinese Academy of Sciences (CAS) “Light of West China” Program, No. 2022-XBQNXZ-013, the Tianchi Talent project, the Natural Science Foundation of Xinjiang Uygur Autonomous Region, No. 2022D01E86, and National Natural Science Foundation of China under grant U2031204.

This research was supported by the Korea Astronomy and Space Science Institute under the R\&D program (Project No. 2024-1-860-02) supervised by the Ministry of Science and ICT.

This work acknowledges funding from the Centre National d'\'Etude  Spatial (CNES) via an AIM/PLATO grant.

The corresponding author, Jayanand Maurya, acknowledges the support of the Physical Research Laboratory at Ahmedabad, India, where the initial work of this paper was carried out.     

This work has made use of data from the European Space Agency (ESA) mission
{\it Gaia} (\url{https://www.cosmos.esa.int/gaia}), processed by the {\it Gaia}
Data Processing and Analysis Consortium (DPAC,
\url{https://www.cosmos.esa.int/web/gaia/dpac/consortium}). Funding for the DPAC has been provided by national institutions, in particular, the institutions
participating in the {\it Gaia} Multilateral Agreement.

This research has made use of the tool provided by \textit{Gaia} DPAC (https://www.cosmos.esa.int/web/gaia/dr3-software-tools) to reproduce (E)DR3 \textit{Gaia} photometric uncertainties described in the GAIA-C5-TN-UB-JMC-031 technical note using data in \citet{2021A&A...649A...3R}.
 
%%%%%%%%%%%%%%%%%%%%%%%%%%%%%%%%%%%%%%%%%%%%%%%%%%
\section*{Data Availability}

 The derived data generated in this study will be shared upon reasonable
request to the corresponding author. All other data used for the present study are publicly available.



\bibliographystyle{mnras}
\bibliography{example} % if your bibtex file is called example.bib

%\begin{thebibliography}{99}
%\end{thebibliography}

%%%%%%%%%%%%%%%%%%%%%%%%%%%%%%%%%%%%%%%%%%%%%%%%%%

%%%%%%%%%%%%%%%%% APPENDICES %%%%%%%%%%%%%%%%%%%%%

%\appendix

%%%%%%%%%%%%%%%%%%%%%%%%%%%%%%%%%%%%%%%%%%%%%%%%%%

% Don't change these lines
\bsp	% typesetting comment
\label{lastpage}
\end{document}